\documentclass[nato,noid,numreferences]{crckbked}

\usepackage{graphicx}
\usepackage{epsfig}
 \newcommand{\zr}[1]{\mbox{\hspace*{#1em}}}

 \newcommand{\RR}{\mbox{\zr{0.1}\rule{0.04em}{1.6ex}\zr{-0.05}{\sf R}}}
 
 \newcommand{\ZZ}{\mbox{\sf Z\zr{-0.45}Z}}

\newcommand{\gl}{\mathrel{\rlap{\lower2pt\hbox{\small \hskip1pt$<$}}
\hspace*{1pt}\raise3pt\hbox{\small $>$}}}

\def\Quadrat#1#2{{\vcenter{\hrule height #2
  \hbox{\vrule width #2 height #1 \kern#1
    \vrule width #2}
  \hrule height #2}}}
\def\dAlemb{\mathop{\kern 1pt\hbox{$\Quadrat{5pt}{0.4pt}$} \kern1pt}}
\def\dAlember{\mathop{\kern 1pt\raise-2pt\hbox{$\Quadrat{3pt}{0.4pt}$} \kern1pt}}

\begin{document}


\begin{opening}
\title{Electric Flux Sectors and Confinement} 



\author{Lorenz von Smekal}
\institute{Institut f\"ur Theoretische Physik III,\\ Universit\"at
     Erlangen-N\"urnberg, D-91058 Erlangen, Germany}

\author{with}
\institute{}

\author{Philippe de Forcrand}
\institute{Institut f\"ur Theoretische Physik, ETH-H\"onggerberg,\\ 
     CH-8093 Z\"urich, Switzerland, and\\ 
     Theory Division, CERN, CH-1211 Gen\`eve 23, Switzerland}

\begin{abstract}
We study the fate of static fundamental charges in the
thermodynamic limit from Monte-Carlo simulations of 
$SU(2)$ with suitable boundary conditions. 
\end{abstract}

\end{opening}

\section{Introduction}

In QED, the charge of a particle is of long-range nature.
It can exist because the photon is massless.
Localized objects are neutral like atoms.
Within the language of local field-systems one derives more generally 
that every gauge-invariant {\em localized} state is singlet 
under the unbroken charges of global gauge invariance.
Thus, without (electric) Higgs mechanism, QED and QCD have in common
that any localized physical state must be chargeless/colorless.

The extension to {\em all} physical states is possible only with a mass gap. 
Without that, in QED, non-local 
charged states which are gauge-invariant can arise as limits of local ones
which are not. The Hilbert space decomposes into the so-called superselection
sectors of the physical states with different charges.
With a mass gap in QCD, on the other hand,  color-electric charge
superselection sectors cannot arise: {\em
every} gauge-invariant state can be approximated by gauge-invariant localized
ones (which are colorless). One concludes that {\em every} gauge-invariant
state must also be a color singlet.

On the other hand, charged states are always possible with suitable boundary
conditions in a finite volume.~This allows to study their fate in the
thermodynamic limit from Monte-Carlo simulations on finite lattices.
In an Abelian theory for example, anti-periodic (spatial) boundary conditions
can be used to force the system into a charged sector in the infinite volume
limit \cite{Pol91}.  
The (Higgs vs. Coulomb) phases of the non-compact Abelian Higgs model can be 
distinguished in this way. And by duality, via the $\ZZ$ gauge theory, 
the magnetic sectors of compact $U(1)$ follow an analogous pattern.
The difference in free energy of the anti-periodic
vs. the periodic ensemble thereby tends to zero or a finite value 
for the (magnetic) Higgs or Coulomb phases, respectively. 

In pure $SU(N)$ gauge theory, one expects the free energy $F_q(T,L)$ of a
static fundamental charge in a $1/T\times L^3$ box, for $L\!\to\!\infty$, to
jump from $\infty$ to a finite value  at $T\!=\!T_c$ reflecting the
deconfinement transition. 
The Polyakov loop $P$ is commonly used to demonstrate this in lattice studies.
If $\langle P \rangle \equiv e^{-F_q/T}\!$, the center 
symmetric (broken) phase gives for $F_q$ an infinite (finite) value. 
However, the periodic boundary conditions (b.c.) within which $\langle  P
\rangle$ is measured are incompatible with the presence of a single charge
also in this case.  
And, like any Wilson loop, $\langle P \rangle$ is subject to UV-divergent
perimeter terms, such that $\langle P \rangle = 0$ at all $T$ 
as the lattice spacing $a \!\rightarrow\! 0$.
 
Following \cite{tHo79}, it is possible, however, to 
measure the gauge-invariant, UV-regular free energy of a static fundamental
charge \cite{deF01,deF01b}, and show that it has the expected behaviour, dual
to that of temporal center flux \cite{Kov00}. 
The preparation of suitable b.c.'s to achieve this is a little indirect.

\section{Twist vs. Electric Flux Sectors in $SU(2)$}

For the different sectors relevant to the confinement transition 
in pure $SU(N)$ gauge theory, one needs to distinguish between the 
finite volume partition functions of two types.  

First, 't~Hooft's twisted boundary conditions fix the total number of
$\ZZ_N$-vortices modulo $N$ that pierce planes of a given orientation. 
On the 4-dimensional torus $T^4$ there are $N^6$ different such sectors
corresponding to the 6 possible orientations for the planes of the twists. 
Without fields that faithfully represent the center $\ZZ_N$ of $SU(N)$, 
the structure is $G=SU(N)/\ZZ_N $ with first homotopy $\pi_1(G) = \ZZ_N$.
The $N^6$ inequivalent choices for imposing (twisted) boundary conditions on
the gauge potentials $A$  
therefore correspond to the classification of the bundles, by their 
$\ZZ_N$-vortex numbers, according to the harmonic 2-forms over 
$T^4$ with $\pi_1(G)=\ZZ_N $ coefficients, the 2nd de~Rahm cohomology
group $H^2(T^4,\ZZ_N)$.

\begin{figure}[t]

\parbox[b]{.6\linewidth}{\epsfig{file=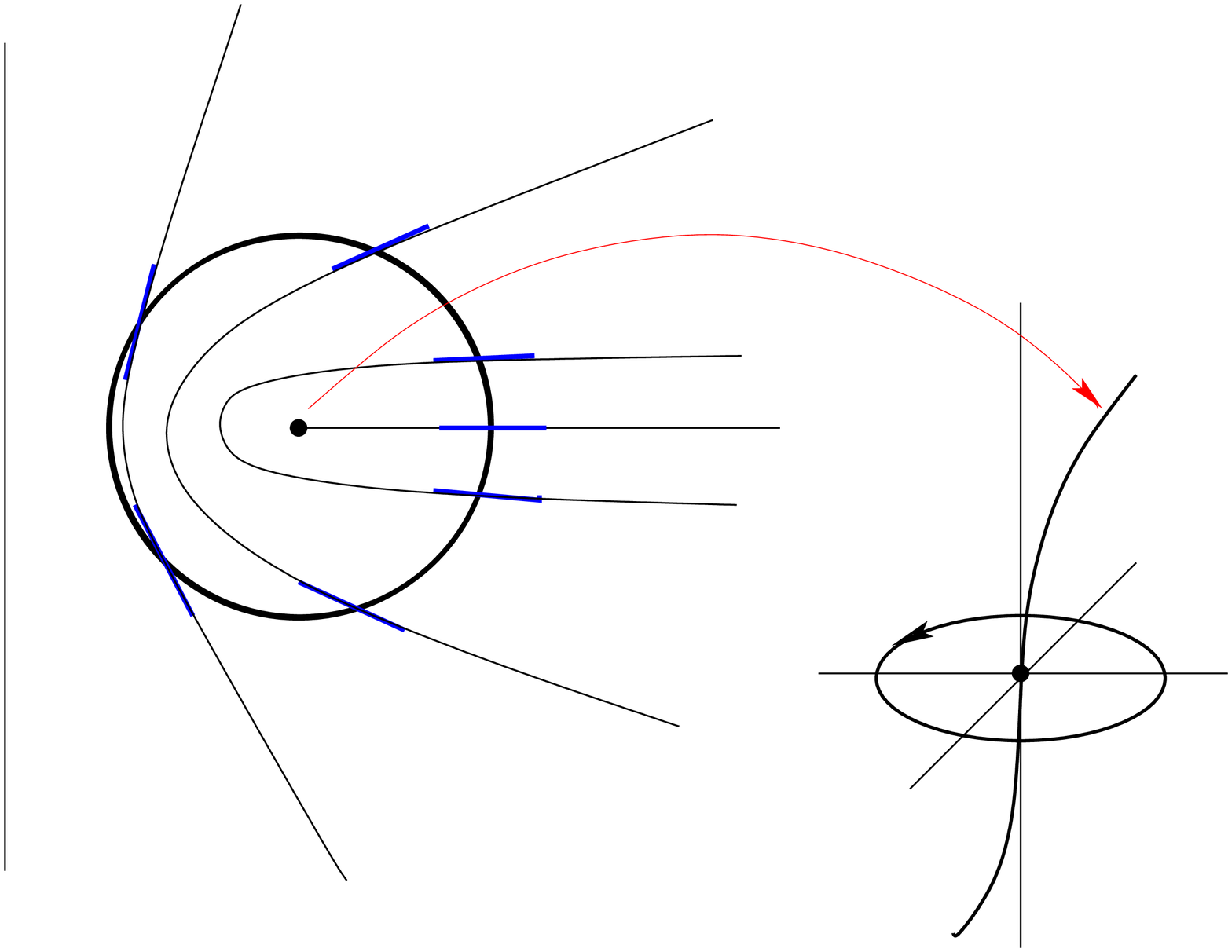,width=\linewidth}

\vspace*{.3cm}

\scriptsize
3d-line defect:~$\ZZ_2$-vortex, maps circle to a non-con\-tracti\-ble
loop in $\RR P(2)$, same happens in $SO(3)\!=\!SU(2)/\ZZ_2$.

\vspace*{.3cm}} \hfill
\parbox[b]{.3\linewidth}{\epsfig{file=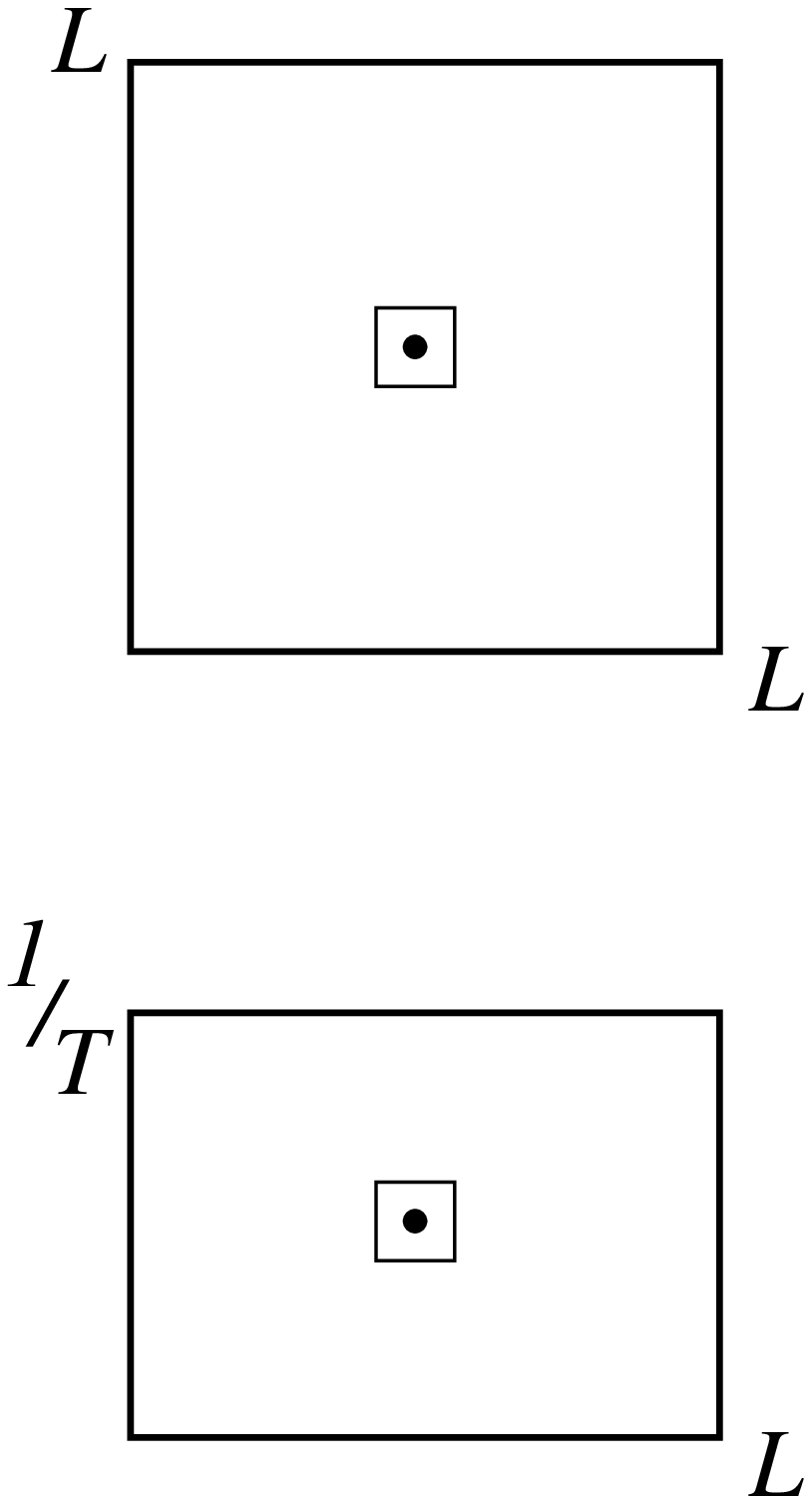,width=\linewidth}}

\caption{3d $\ZZ_2$-vortex as in, {\it e.g.}, nematic liquid  
crystals (left). Assume the 4d-vortices can lower their free energy by
spreading (right): they can do that in the 3 spatial $\vec m$-planes 
at all $T$ (top), 
while they are squeezed in the 3 temporal $\vec k$-planes 
at large $T$ (bottom).}\label{Fig1}
\end{figure}

At finite temperature $T>0$ the $N^6$ possible twists come in two classes: 
3 temporal ones classified by a vector $\vec k \in \ZZ_N^3$, and 3 
magnetic ones by $\vec m \in \ZZ_N^3$, see Fig.~\ref{Fig1}. 
Magnetic twist is defined in purely spatial planes and fixes the conserved,
$\ZZ_N$-valued and gauge-invariant magnetic flux $\vec m$ in the
perpendicular directions.

The different choices of twisted b.c.'s lead to sectors of fractional 
Chern-Simons number ($\nu + \vec k\cdot\vec m /N$) \cite{vBa82} 
with states labelled by $|\vec k , \vec m , \nu \rangle $, where $\nu
\in \ZZ$ is the usual instanton winding number.
These sectors are connected by homotopically non-trivial
gauge transformations $\Omega [\vec k,\nu]$, 
\begin{equation}
\label{hom-nt}
      \Omega [\vec k',\nu'] \, |\vec k , \vec m , \nu \rangle \, =\,  
          |\vec k\! +\!\vec k', \vec m , \nu\! +\!\nu' \rangle \, .
\end{equation}
A Fourier transform of the twist sectors $Z_k(\vec k, \vec m, \nu)$, which
generalizes the construction of $\theta$-vacua as 
Bloch waves from $\nu$-vacua in two ways, 
by replacing $\nu \to (\nu + \vec k\cdot\vec m /N)$ 
for fractional winding numbers and with an additional
$\ZZ_N^3$-Fourier transform w.r.t.~the temporal twist $\vec k$, yields, 
\begin{equation}
 Z_e(\vec e,\vec m,\theta )\,=\, 
   e^{-\frac{1}{T} F_e(\vec e, \vec m,\theta)}  
        \,=\, \frac{1}{N^3} \, \sum_{\vec
 k, \,\nu} \,  e^{-i\omega(\vec k, \nu)} \,
 Z_k(\vec k,  \vec m, \nu)  \, .  \label{ZnFT}
\end{equation}    
Up to a geometric phase $ \omega(\vec k,\nu) =
2\pi \vec e\cdot\vec k/N  + \theta (\nu \!+\! \vec k\cdot\vec m /N)$,  
the states in the new sectors are then invariant under the non-trivial
$\Omega[\vec k,\nu]$ also, 
\begin{equation}
\label{hom-nt-inv}
      \Omega [\vec k,\nu] \, |\vec e , \vec m , \theta \rangle \, =\,  
         \exp\{ i \omega(\vec k,\nu) \} \, 
  |\vec e , \vec m , \theta \rangle \, .
\end{equation}
Their partition functions $Z_e$ are classified, in addition to their magnetic
flux $\vec m $ and vacuum angle $\theta $, by their $\ZZ_N$-valued  
{\em gauge-invariant electric flux} in the $\vec e$-direction \cite{tHo79}.
Here, we do not consider finite $\theta$ which we omit henceforth. 
Recall the following points for details of which we refer to
\cite{deF01}:

{\bf (i)} The twisted partition functions, relative to the periodic ensemble
$Z_k(0,0)$, are expectation values of combinations of 't~Hooft loops
$\widetilde W$ of maximal size in $(\mu,\nu)$-planes dual to the planes of
the twists, 
\begin{equation}
\label{eq:4}
     {Z_k(\vec k, \vec m)}/{Z_k(0,0)} =
              \langle{\widetilde{W}}^{\rm  max}_{(\mu,\nu)}\, \rangle \; .
\end{equation}
In particular, the temporal $\vec k$-twists correspond to expectation values  
of spatial 't~Hooft loops. The $\ZZ_N^3$-Fourier transform of Eq.~(\ref{ZnFT}) 
exhibits their Kramers-Wannier duality with the electric flux sectors which
are expectation values of Polyakov loops in the no-flux ensemble $Z_e(0,0)$
(see below).   

\begin{figure}[t]
  \centering{\
        \epsfig{file=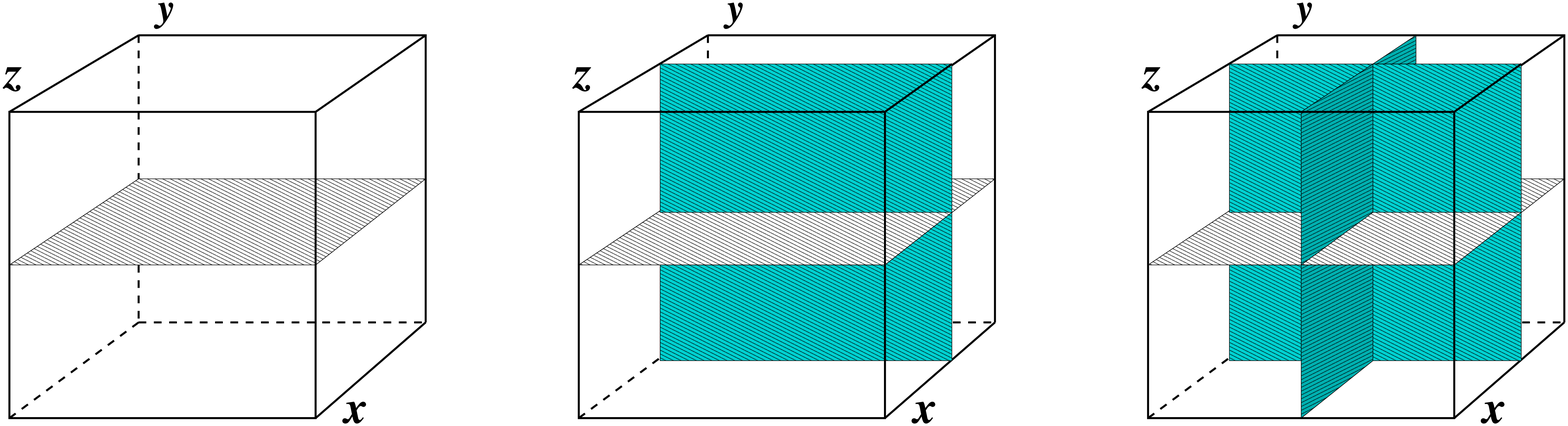,width=.85\linewidth}}
\vspace{.2cm}
\caption{Cubes with one, two, and three $L\!\times\! L$ planes
dual to the stacks of plaquettes flipped for temporal 
twist $\vec k = (0,0,1)$, $\vec k = (0,1,1)$, and
$\vec k = (1,1,1)$, from left to right.}  
\label{tHloops} 
\end{figure}

{\bf (ii)} Note also that the no-flux ensemble in a finite volume is
manifestly different from the periodic ensemble, {\it e.g.}, for
$SU(2)$ one has,
\begin{equation}
\label{eq:5}
  Z_e(0,0) = \frac{Z_k(0,0)}{8}
 \Big(1+3 \langle {\widetilde{W}}^{\rm  max}_{1,\rm spat} \, \rangle + 
3\langle {\widetilde{W}}^{\rm  max}_{2,\rm spat} \, \rangle + 
     \langle {\widetilde{W}}^{\rm max}_{3,\rm spat} \, \rangle \Big)
\end{equation} 
The combinations of spatial 't~Hooft loops needed to compute this, or any of 
the electric flux partition functions $Z_e(\vec e,0) $, are sketched in
Fig.~\ref{tHloops}. 


\begin{figure}[t]
\vspace*{-.5cm}

\hfill
\parbox[b]{.48\linewidth}{\rightline{\epsfig{file=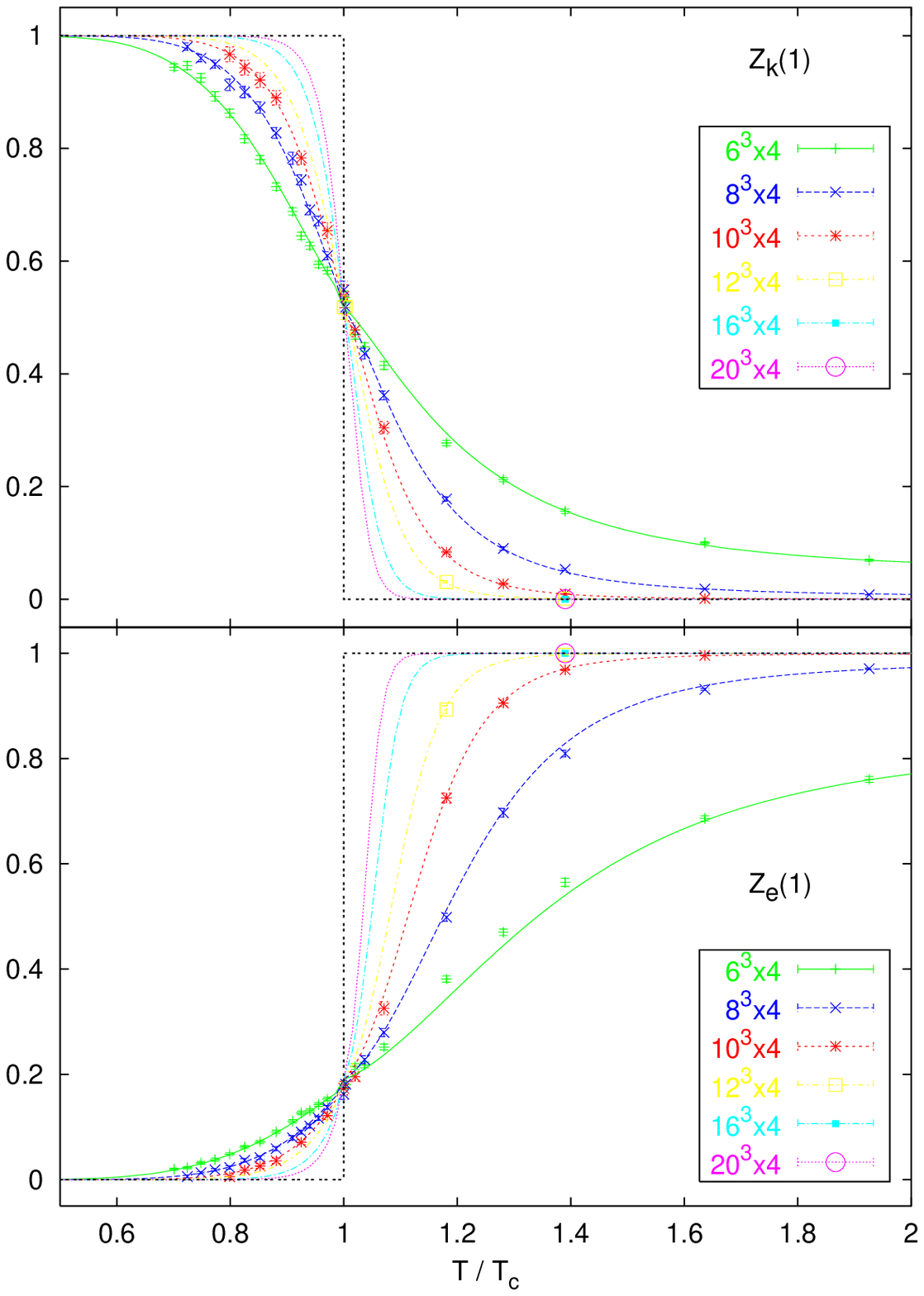,width=1.092\linewidth}}}
\hskip .1cm
\parbox[b]{.48\linewidth}{\rightline{\epsfig{file=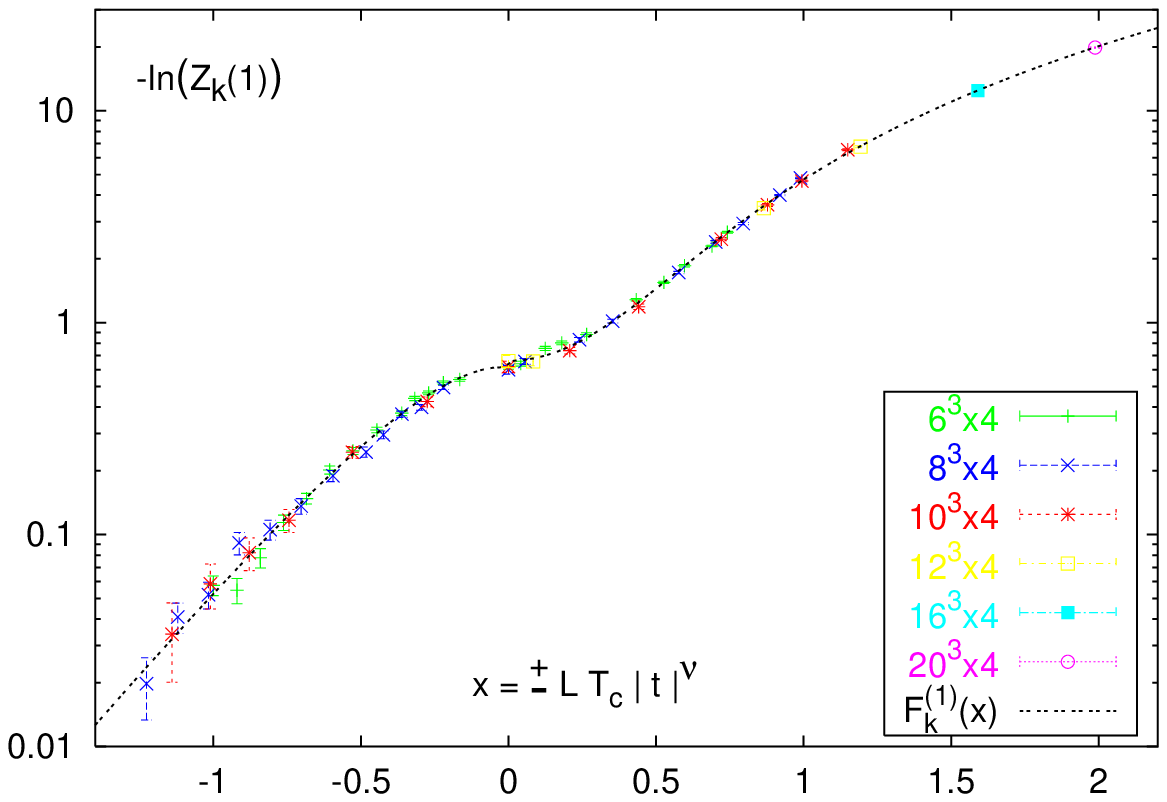,width=1.06\linewidth}}
\vskip -.4cm
\epsfig{file=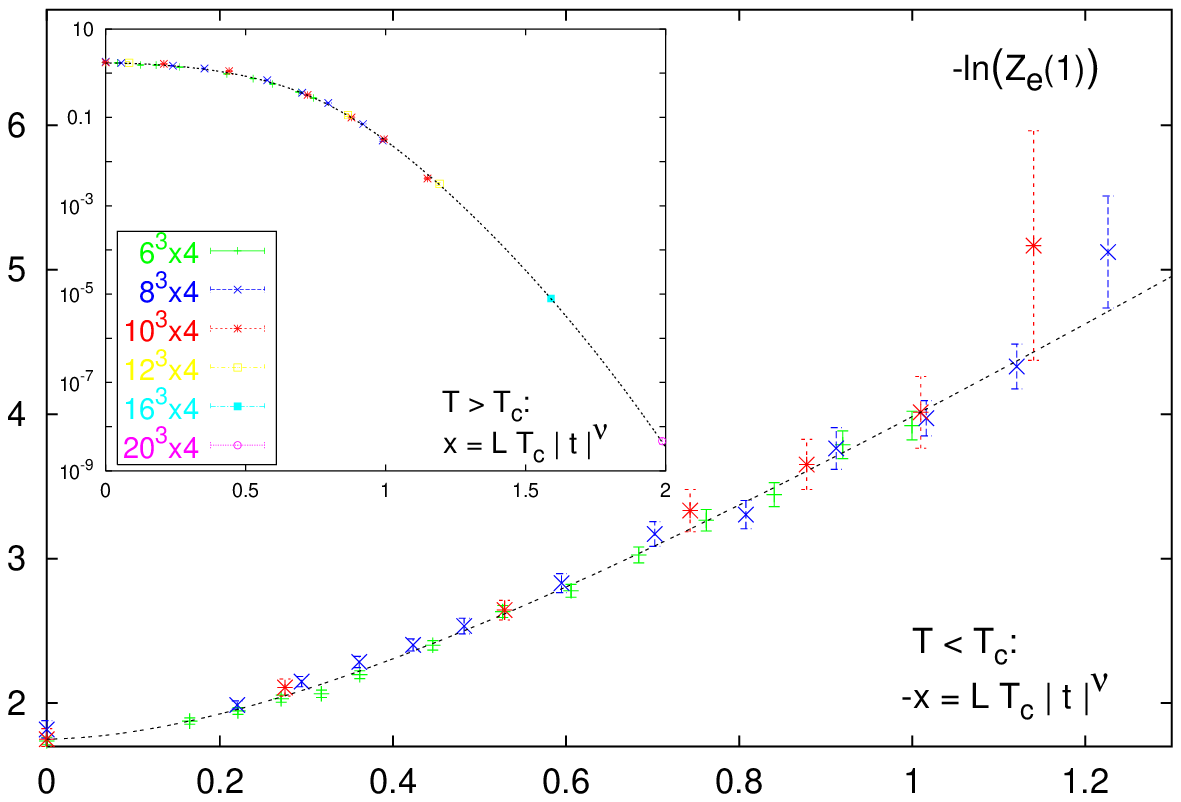,width=\linewidth}} 
\caption{Partition function and free energy of temporal twist (top) and
  electric flux (bottom), both over 
  temperature (left) and finite size scaling variable $x$ (right).}\label{Fig3}
\end{figure}

{\bf (iii)} From the gauge-invariant definition of the 
Polyakov loop  $P(\vec x)$  in presence of temporal twist \cite{vBa84}, 
it is relatively simple but important to verify
that the electric-flux partition functions are indeed expectation
values of $P$'s in the no-flux ensemble \cite{deF01}, 
which follow the general pattern, 
\begin{equation}
\label{eq:6}
     \frac{Z_e(\vec e, 0)}{Z_e(0,0)} =  
       \big\langle P(\vec x) P^\dagger(\vec x +L\vec e) \big\rangle _{L,T}
  \stackrel{L\to\infty}{\longrightarrow}  
  \left\{ { 
    0 \, , \;\;
\langle{\widetilde{W}}^{\rm  max}_{i,\rm spat}\,\rangle  \to 1 \, , \; T < T_c
                                                   \atop 
      1 \, , \;\;
\langle {\widetilde{W}}^{\rm  max}_{i,\rm spat}\,\rangle \to 0 \, , \; T > T_c
                            } \right.
\end{equation}  
with $\vec e \in \ZZ_N^3$ again. For $\vec e \not=0$ this therefore yields 
the free energy $ F_q(T,L) = F_e(\vec e,0)$ of one static fundamental charge 
in the volume $L^3$ with b.c.'s such that its electric flux is directed 
towards its `mirror' (anti)charge in the adjacent volume along the direction
of $\vec e$. Also note that the operator in the expectation 
value of (\ref{eq:6}) has no perimeter, 
is UV-regular, and one can see in Fig.~\ref{Fig3} that there is no Coulomb
term for small volumes either.      

Of course, Eq.(\ref{eq:6}) reflects the different realizations of the
electric $\ZZ_N^3$ center symmetry in the respective phases. 
As compared to spin correlations of the form $\langle s_{\vec x} \,s_{\vec x
+L\vec  e_i}\rangle $ in the 3d-Ising model with interfaces, the Polyakov
loops in (\ref{eq:6}) are the corresponding variables in $SU(2)$, whose
behavior as a function of temperature is reversed. 

The dual of the Ising model on the other hand is the 3d $\ZZ_2$ gauge theory.
Interfaces in the first are Wilson loops in the latter. 
Through duality, the expectation value of a $\ZZ_2$-Wilson loop along 
$C$ is expressed as ratio of Ising-model partition functions with and
without antiferromagnetic bonds at links dual to a surface
$\Omega$, the $\ZZ_2$-interface, bounded by $C$, {\it e.g.}, see
\cite{Sav80},   
\begin{equation}
\label{eq:7}
\big\langle W_{\ZZ_2}(C) \big\rangle = \frac{\sum_{\{s\}}
  \exp\sum_{<ij>} \beta_{ij} \, s_is_j }{\sum_{\{s\}}  \exp\beta\sum_{<ij>}
  s_is_j  } \; , \;\; 
 \beta_{ij}  = \Bigg\{  { - \beta \; , \;\;  <\!ij\!>
    \in \Omega^* \atop \phantom{-}  \beta \; , \;\; <\!ij\!> 
      \not\in \Omega^* }  \; .
\end{equation}
In $SU(2)$, the objects dual to the  Polyakov loop correlations, or the
electric fluxes in (\ref{eq:6}), are spatial 't~Hooft loops. Via universality,
these are the $\ZZ_2$-Wilson loop analogues. And their expectation values are
calculated on the lattice in much the same way, by flipping a coclosed set 
$\Omega^*$ of plaquettes dual to some surface $\Omega $ subtended by the 
loop $C$,  
\begin{equation}
\label{eq:8}
\big\langle \widetilde{W}_{SU(2)}(C) \big\rangle = \frac{\int
  dU \exp -\sum_{\dAlember } \beta( \dAlemb ) \mbox{Tr}\, U\!_{\dAlember} }
     {\int dU   \exp -\beta \sum_{\dAlember } \mbox{Tr}\, U\!_{\dAlember} }
   \; , \;\; 
 \beta(\dAlemb )  = \Bigg\{  { - \beta \; ,
  \;\;  
  \dAlemb \in \Omega^* \atop \phantom{-}  \beta \; , \;\; \dAlemb \not\in
          \Omega^* } \; .
\end{equation}
In both cases the surface is arbitrary except for its
boundary $C=\partial\Omega$.

\pagebreak

\noindent
A spatial 't~Hooft loop of maximal size $L\times L $, living in, say, the
$(x,y)$ plane of the dual lattice, is equivalent to
an odd number of flipped plaquettes in every $(z,t)$ plane of the original
lattice. This enforces twisted b.c.'s in $(z,t)$. 
Combining such loops yields the other $\vec k$-twist sectors, {\it
c.f.}, Fig.~\ref{tHloops}. 

The temperature dependences of the partition functions for temporal twist 
$\vec k \!=\! (0,0,1)$ and electric flux $\vec e \!=\!(0,0,1) $ 
as calculated in \cite{deF01} are compared in Fig.~\ref{Fig3}. Their dual
behavior is obvious: for $L\!\to\!\infty$, both  
approach step functions jumping from 1 to 0 and 0 to 1, respectively, 
as $T_c$ is crossed (from below). Near the phase transition, this behavior is
determined by critical exponents and likewise universal amplitude
ratios of the 3d-Ising class. For the larger spatial lattice sizes, the 
fits in the left part of Fig.~\ref{Fig3} might look rather daring at
first. 
However, each of the two families of curves shown there
really represent one of the unique functions in the right part
which fit all the data. This is possible due to finite size scaling.

\section{Finite-Size Scaling}

\begin{figure}[t]
\parbox[b]{.48\linewidth}{\rightline{\epsfig{file=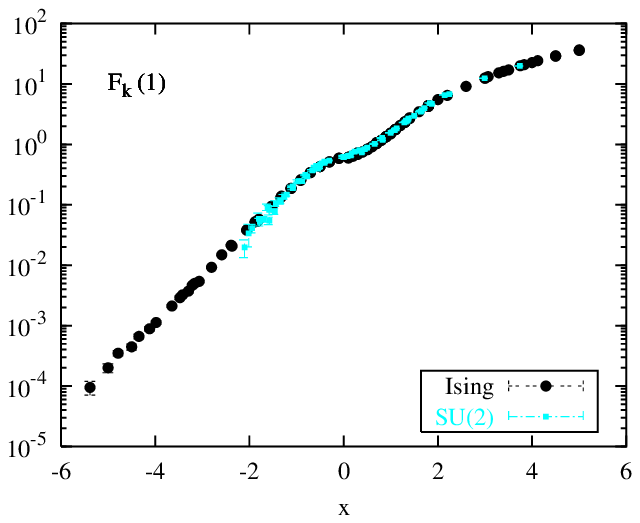,width=1.14\linewidth}}}
\hskip -.2cm
\parbox[b]{.48\linewidth}{\leftline{\epsfig{file=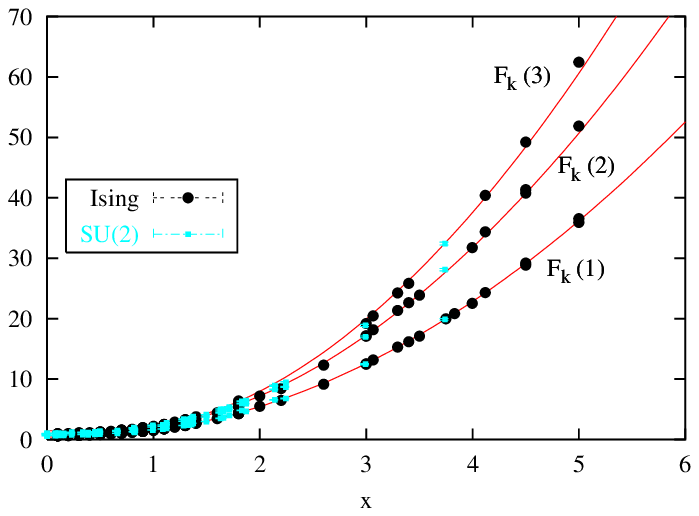,width=1.14\linewidth}}}
\caption{Free energy $F_k(1)=-\ln Z_k(1)$ of one temporal twist in $SU(2)$,
             same data as in Fig.~\protect\ref{Fig3}, compared to  
             that of an interface in the Ising model 
             $F_k (1) = -\ln(Z_a/Z_p)$ with $x_{\mbox{\tiny Ising}} = -1.88(2)
             x_{SU(2)}$ (left);
             1-3 orthogonal $\vec k$-twists in
             $SU(2)$ ($T\!>\!T_c$)  vs. Ising with 1-3 anti-periodic 
             directions ($T\!<\!T_c$), with square-root ratios in the
             fits (right), from   \protect\cite{deF01c}.}\label{Fig4} 
\end{figure}

Finite size scaling (FSS) laws are based on the observation that the length 
of the system $L$ and the correlation lengths $\xi$ that diverge in the
thermodynamic limit are the only relevant length scales in the neighborhood
of the transition. In particular, as the critical point is approached, the
finite lattice spacing $a$ becomes less and less important.    
For the continuous (2nd order) transition of $SU(2)$ 
in the 3d-Ising class, with $\xi_{\pm}(t) = \xi^0_\pm  |t|^\nu$ for reduced
temperature $t \gl 0$, we use $ t = T/T_c - 1 $ and $\nu = 0.63$. 
And as for the ratios of 3d-Ising model partition
functions $Z_a/Z_p$ with anti-periodic vs. periodic b.c.'s \cite{Has93}, we
assume  the $L$, $T$ dependence of the various temporal twist sectors,
denoting $ Z_k(i) = \langle {\widetilde{W}}^{\rm  max}_{i,\rm spat}\,\rangle
$ for $i = 1,..3$ orthogonal $\vec k$-twists, to be governed by simple FSS
laws, 
\begin{equation}
\label{eq:9}
  \hskip 1cm Z_k(i) \, =  \, f^{(i)}(x) \; , \;\;  \mbox{with} 
\;\;  x = \pm L T_c |t|^\nu  \propto  L/\xi_{\pm}(t)      \; , \;
 \mbox{for} \;  t \gl  0 \;   . \label{fss} 
\end{equation}
We then observe that our results over $x$ for all different lattice sizes
nicely collapse on a single curve, {\it c.f.}, Fig.~\ref{Fig3}.
In the high temperature phase above $T_c$, the large-$x$ behavior, $ -\ln
f^{\scriptscriptstyle (i)}(x) =  \tilde\sigma_0^{\scriptscriptstyle (i)}\,
x^2 + \dots $, reflects the dual area 
law for ($i=1,.. 3$) large spatial 't~Hooft loops. Their dual string tension, 
\begin{equation}
\label{eq:10}
     \tilde\sigma(t) = \tilde\sigma_0^{\scriptscriptstyle (1)} T_c^2
     \; t^{2\nu} \, =
       \,    R/\xi_+^2(t) \;  ,
\end{equation}
is the Ising analogue of the interface tension $ \sigma_I(t) =
R/\xi_-^2(t) \sim |t|^{2\nu} $ below $T_c$, where $R\simeq 0.104$ is a
universal ratio \cite{Kle92,Has97}. 
In addition, the universality hypothesis relates the ratio 
of the correlation lengths for the Polyakov loops in $SU(2)$ to 
the correlation lengths of the spins in the 3d-Ising model, as measured in
\cite{Has97}, \vspace*{-.2cm} 
\begin{equation}
\label{uar}
 \xi_-^{SU(2)}/\xi_+^{SU(2)}  \,  \stackrel{!}{=} \, 
    \xi_+^{\mbox{\tiny Ising}}/\xi_-^{\mbox{\tiny Ising}} \,  \simeq \, 1.96
    \; .
\end{equation}    
Together with the large-$x$ asymptotics of $i= 1,..3$ 
electric fluxes, $ - \!\ln Z_e(i)= \sigma_0^{\scriptscriptstyle (i)} |x| +
\dots $, below $T_c$, this relates the string tension amplitude $\sigma_0 
\equiv \sigma_0^{\scriptscriptstyle (1)} $ 
below $T_c$ to its dual counterpart
$\tilde\sigma_0\equiv\tilde\sigma_0^{\scriptscriptstyle (1)}$ 
above $T_c$, 
\begin{equation}
\label{eq:12}
 \sigma(T) = \sigma_0\,  T_c \, |t|^\nu \, T \; , \;\; \mbox{with} \;\;
             \sigma_0 =
   \sqrt{\tilde\sigma_0/R_+}
       \; , \;\;   R_+ =  R \, {\xi_-^2}/{\xi_+^2}\,  \simeq 0.4 \; .
\end{equation}
Within the accuracy of our results, the $SU(2)$ data is fully consistent with
these ratios. The quite impressive range of universal behavior can be
appreciated in comparing the $SU(2)$ temporal twist with the interface
free energy $-\ln Z_a/Z_p$ in the 3d-Ising model \cite{deF01c}, 
see Fig.~\ref{Fig4}. Once a non-universal constant of proportionality 
relating both FSS variables is fixed, the scaling functions 
appear to be identical for the whole range of the $SU(2)$ data (with 
high and low temperature phases interchanged). The same agreement is observed
for all $f^{(i)}(x)$, {\it i.e.},  also  between 2 and 3 orthogonal twists
in $SU(2)$ and Ising model with anti-periodic b.c.'s in 2 and 3 directions.

\section{String Formation}

In the low temperature phase, the formation of electric flux strings is
expected. The signature for this are square-root ratios $ 1 : \sqrt{2} :
\sqrt{3} $ of the string tension amplitudes $\sigma_0^{\scriptscriptstyle (i)}
$ for $i=1, 2 $  and 3 orthogonal fluxes. At $T=0$ the
singificance for such a behavior, as compared to the ratios $ 1 : {2} :
{3} $ expected for isotropic fluxes, was assessed in the pioneering study of
Ref.~\cite{Has90}.   

Above $T_c$ on the other hand, the same square-root ratios
for the dual string tension amplitudes, 
\begin{equation}
 \tilde\sigma_0^{(1)} : \tilde\sigma_0^{(2)} : \tilde\sigma_0^{(3)} 
      \; \sim \;  1 : \sqrt{2}:\sqrt{3}  \; ,   
\end{equation}
signal the formation of interfaces with minimal area. These ratios 
are well confirmed for spatial 't~Hooft loops in orthogonal planes within 
the accuracy of our $SU(2)$ data \cite{deF01}, and with considerably higher
accuracy also for the 3d-Ising model (for $T<T_c$) with anti-periodic b.c.'s
in 1,2 and 3 directions \cite{deF01c}, which are also compared in
Fig.~\ref{Fig4}.   

\section{Conclusions}

To summarize, we have studied the finite volume partition
functions in the sectors of pure $SU(2)$ of two types:
Using 't Hooft's twisted boundary conditions we  
first measured the free energies of ensembles with odd
numbers of $\ZZ_2$ center-vortices through temporal planes.  
From combinations of these we then obtained the sectors of
gauge-invariant electric flux, and 
demonstrated explicitly that, below $T_c$, their 
free energy diverges linearly with the length $L$ of the system. 
Because spatial twists share their qualitative low-$T$ behavior with 
the temporal ones considered so far, the free energy of the magnetic fluxes
must vanish just as that of temporal twist. 
This is the magnetic Higgs phase with electric confinement 
of pure $SU(2)$.

At criticality all free energies rapidly approach their finite 
$L\to\infty$ limits. 
Above $T_c$, the electric-flux free energies vanish in the
thermodynamic limit. The transition is well described by simple 
finite size scaling laws of the 3d-Ising class. 
The ratios of the (dual) string tension amplitudes 
for 1,2 and 3 orthogonal (spatial 't~Hooft loops) electric fluxes 
(above) below $T_c$ indicate the formation of diagonal (interfaces) 
flux strings.


Both of us would like to express our warm thanks 
to \v{S}.~Olejn{\'i}k and J.~Greensite for their great job
organizing this stimulating workshop.


\end{document}